\documentclass{nature}
\usepackage{graphicx}

\makeatletter
\let\saved@includegraphics\includegraphics
\AtBeginDocument{\let\includegraphics\saved@includegraphics}
\renewenvironment*{figure}{\@float{figure}}{\end@float}
\makeatother

\usepackage{dcolumn}
\usepackage{bm}
\usepackage{xspace}
\usepackage{wrapfig}
\usepackage{amsmath,amsfonts,amssymb,bm,ulem}
\usepackage[usenames]{color}

\renewcommand{\vec}[1]{{\ensuremath{\bm{\mathrm{#1}}}}}

\newcommand{\YIG}{YIG\xspace}
\newcommand{\CrO}{Cr$_{2}$O$_{3}$\xspace}

\title{Spin colossal magnetoresistance in an antiferromagnetic insulator}

\author{Zhiyong Qiu$^{1,2}$, Dazhi Hou$^{3 \star}$, Joseph  Barker$^{1}$, Kei Yamamoto$^{1,4,5,6}$, Olena Gomonay$^{4}$, Eiji Saitoh$^{1,3,6,7}$}

\begin{document}

\maketitle

\begin{affiliations}
        \item Institute for Materials Research, Tohoku University, Sendai 980-8577, Japan
        \item Key Laboratory of Materials Modification by Laser, Ion, and Electron Beams (Ministry of Education), School of Materials Science and Engineering, Dalian University of Technology, Dalian 116024, China
        \item WPI Advanced Institute for Materials Research, Tohoku University, Sendai 980-8577, Japan
        \item Institut f\"ur Physik, Johannes Gutenberg Universit\"at Mainz, D-55128, Mainz
        \item Department of Physics and Astronomy, The University of Alabama, AL 35487, USA and Centre of Materials for Information Technology, The University of Alabama, AL 35401, USA
        \item Advanced Science Research Center, Japan Atomic Energy Agency, Tokai 319-1195, Japan
        \item (Present address) Department of applied physics, The University of Tokyo, Tokyo 113-8656, Japan
        
\end{affiliations}

\newpage

\begin{abstract}
	Colossal magnetoresistance (CMR) refers to a large change in electrical conductivity induced by a magnetic field in the vicinity of a metal-insulator transition and has inspired extensive studies for decades\cite{Ramirez1997, Tokura2006}. Here we demonstrate an analogous spin effect near the N\'eel temperature $T_{\rm{N}}$=296 K of the antiferromagnetic insulator \CrO. Using a yttrium iron garnet \YIG/\CrO/Pt trilayer, we injected a spin current from the YIG into the \CrO layer, and collected via the inverse spin Hall effect the signal transmitted in the heavy metal Pt. We observed a change by two orders of magnitude in the transmitted spin current within 14 K of the N\'eel temperature. This transition between spin conducting and nonconducting states could be also modulated by a magnetic field in isothermal conditions. This effect, that we term spin colossal magnetoresistance (SCMR), has the potential to simplify the design of fundamental spintronics components, for instance enabling the realization of spin current switches or spin-current based memories.
\end{abstract}

\begin{figure}
        \centering
        \includegraphics[width=1\linewidth]{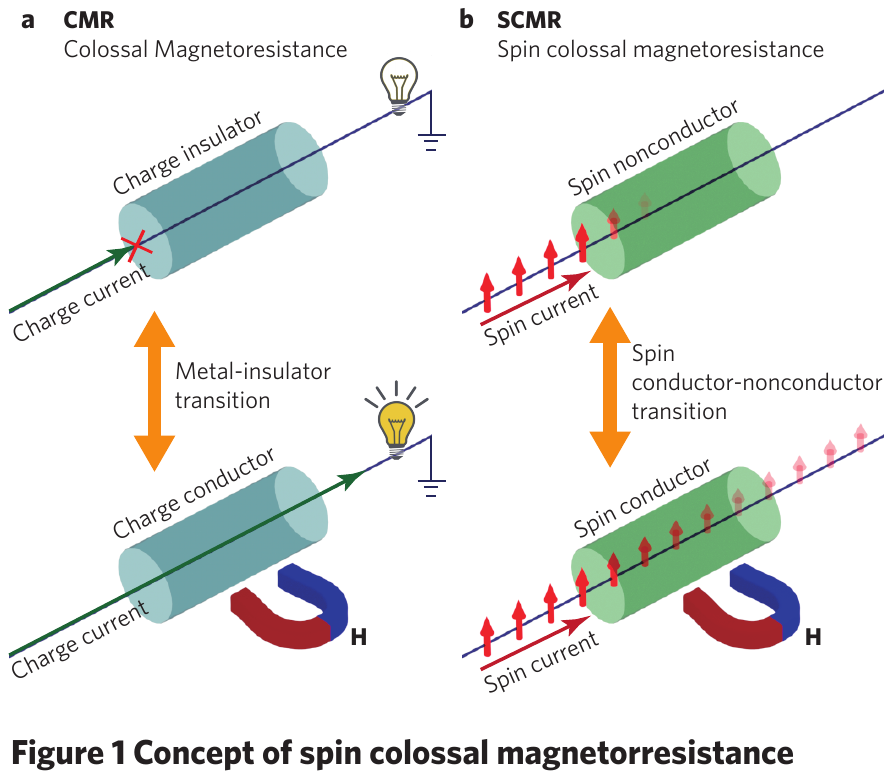}
        \caption{\textbf{Concept of spin colossal magnetoresistance.} 
                \textbf{a}. A schematic illustration of colossal magnetoresistance (CMR). CMR is a property of some materials in which their electrical resistance changes steeply in the presence of a magnetic field, typically due to the strong coupling between a steep metal-insulator transition and a magnetic phase transition.   
                \textbf{b}. A schematic illustration of spin colossal magnetoresistance (SCMR): spin current transmissivity changes steeply due to the change in symmetry (in this paper due to a magnetic phase transition). The spin current transmissivity is also modulated by an applied magnetic field.}
        \label{fig01}
\end{figure}

Spin current is a flow of spin angular momentum sharing many analogues with electric currents. Spin currents can be carried not only by migrating electrons, but also by magnetic quasi-particles such as magnons or spin waves\cite{Zutic2004, Maekawa2006}. These magnetic excitations are of particular interest in the spintronics community because they allow a spin current to flow in electrical insulators where charge currents cannot\cite{Kajiwara2010,Uchida-Nature-2008,Takei-PhysRevB-2014,Cornelissen-NatPhys-2015,Qiu-NatComms-2016}. Thus, insulator spintronics may provide a route towards low power devices where spin currents carry signals and encode information. However, it is not straightforward to create all spintronic components which are analogous to their electronic counterparts\cite{Tserkovnyak-NatNano-2013}. For example, a  basic  spin-current on-off switch has not been demonstrated in insulator spintronics. The difficulty lies in the lack of a simple mechanism to directly gate the carrier density in magnetic insulators. Nevertheless, some clues may be obtained from colossal magnetoresistance (CMR) occurring in materials which exhibit metal-insulator transitions, where a large modulation of charge conductivity can be induced by a magnetic field\cite{Ramirez1997, Tokura2006} (Fig.\ref{fig01} 1a). Spin conductivity may be tunable in systems with a spin conducting-nonconducting transition (Fig.\ref{fig01} 1b).

In this Letter, we report such a `conductor-nonconductor transition' for spin currents in the uniaxial antiferromagnetic insulator \CrO \cite{Brockhouse-JChemPhys-1953,Foner-PhysRev-1963,nagamiyaAFM}. We found that \CrO does not conduct spin currents below the N\'eel temperature, but abruptly becomes a good spin conductor above this temperature. Furthermore, in the vicinity of the transition, the spin-current transmission can be modulated by up to 500\% with a $2.5$~T\ magnetic field: spin colossal magnetoresistance (SCMR). The active transport element is spin angular momentum rather than electrical charge.

\begin{figure}
        \centering
        \includegraphics[width=1\linewidth]{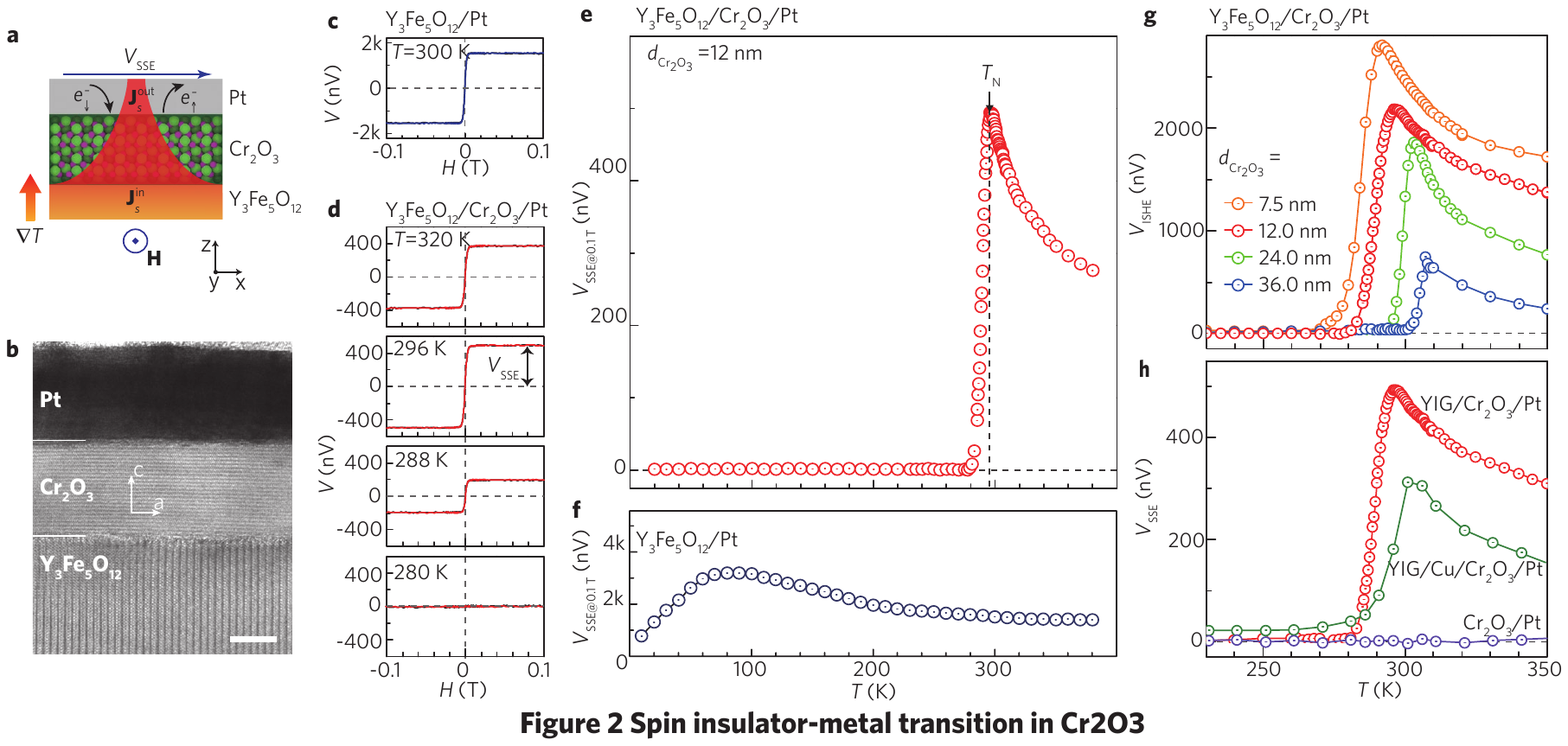}
        \caption{\textbf{Spin conductor-nonconductor transition in \CrO.}
                \textbf{a}. \textcolor{black}{A schematic illustration shows the concept of the spin-current transmissivity measurement of a \YIG/\CrO/Pt trilayer device.}  A temperature gradient, $\nabla T$, is along the $z$ direction while an external magnetic field ($\vec{H}$) is along the $y$ direction. \textcolor{black}{The magnetic insulator \YIG is used as a spin source to inject spin currents $\vec{J}_s^{\rm in}$ into the \CrO based on the spin Seebeck effect, and transmitted spin currents $\vec{J}_s^{\rm out}$ through the \CrO are detected as voltage signals in the Pt layer via the inverse spin Hall effect.}
                \textbf{b}. A cross sectional TEM image of the \YIG/\CrO/Pt trilayer device used in this work. The scale bar of the image is 5 nm. The easy axis $c$ of the \CrO is in the out-of-plane direction $z$ of the film \textcolor{black}{as inserted axis shows.}
                \textbf{c}. The external magnetic field $\vec{H}$ dependencies of the voltage signal $V$ measured in a \YIG/Pt bilayer device at 300 K.
                \textbf{d}. The external magnetic field $\vec{H}$ dependencies of the voltage signal $V$ measured in the \YIG/\CrO/Pt trilayer device at various temperatures.
                \textbf{e}. The temperature dependence of the spin Seebeck voltage $V_{\rm{SSE}}$ at $H=$0.1 T for the \YIG/\CrO/Pt trilayer device.  
                \textbf{f}. The temperature dependence of the spin Seebeck voltage $V_{\rm{SSE}}$ at $H=$0.1 T for a \YIG/Pt bilayer device.
                \textbf{g}. The temperature dependence of spin pumping signals $V_{\rm{ISHE}}$ for YIG/Cr$_{2}$O$_{3}$/Pt trilayer devices with various values of the Cr$_{2}$O$_{3}$ layer thickness $d_{\rm{Cr_{2}O_{3}}}$. 
                \textbf{h}. The temperature dependence of the spin Seebeck voltage $V_{\rm{SSE}}$ at $H=$0.1 T for Y$_3$Fe$_5$O$_{12}$/Cu/Cr$_{2}$O$_{3}$/Pt and Cr$_{2}$O$_{3}$/Pt devices. 
                \textcolor{black}{The measurement errors are smaller than the size of data points in all figures. }
                \label{fig:temp_dep}}     
\end{figure}

The spin-current transmission in \CrO was studied by using a trilayer device, sandwiching a \CrO thin film between a magnetic insulator yttrium iron garnet (\YIG) and a heavy metal Pt layer (Fig. \ref{fig:temp_dep}a). Here, YIG serves as a spin-current source. By using a temperature gradient, $\nabla T$, along the out-of-plane direction $z$, the spin Seebeck effect (SSE)\cite{Uchida-Nature-2008,Xiao-PhysRevB-2010} generates a spin accumulation at the interface of \YIG/\CrO, which drives a spin current ($\vec{J}_s^{\rm in}$) into the \CrO layer. Spin currents, transmitted through the \CrO layer to the Pt interface (${\vec{J}}_s^{\rm{out}}$), are converted into a measurable voltage via the inverse spin Hall effect (ISHE)\cite{Saitoh2006}.     

We define the spin-current transmissivity $\mathcal{T}_{\rm{s}}=|{\vec{J}}_s^{\rm{out}}|/|{\vec{J}}_s^{\rm{in}}|$, describing the relative amount of spin-current incident on the \YIG/\CrO interface which is transmitted to the \CrO/Pt interface. \textcolor{black}{We ignore the effect of the spin accumulation at the \YIG/\CrO interface on ${\vec{J}}_s^{\rm{in, out}}$, which will a $posteriori$ be justified as the data are well explained solely by the intrinsic properties of the \CrO.} The spin current entering the \CrO, $\vec{J}_s^{\rm in}$, flows along $\nabla T$ and is polarized along $\vec{M}$, where $\vec{M}$ is the magnetization of the \YIG layer which can be easily manipulated by the external magnetic field $\vec{H}$. The ISHE voltage measured in the Pt layer is thus
\begin{equation}
V_{\rm{SSE}} \propto  \vec{J}_s^{ \rm{out}}=\mathcal{T}_{\rm{s}}  \left(  \vec{J}_s^{ \rm{in}}\times\frac { \vec{M} }{\left|  \vec{M}\right|} \right) \cdot \hat{\vec{x}} ,
\label{eq1}
\end{equation}
where $\hat{\vec{x}}$ is the unit vector along the $x$ axis. $\vec{J}_s^{\rm in}$ can be roughly estimated by the SSE signal in the \YIG/Pt device in which the \CrO thickness is zero and the SSE signal in \YIG/\CrO/Pt devices yields ${\vec{J}}_s^{\rm{out}}$ \cite{Wang2015a,Moriyama2015}. Therefore, the spin-current transmissivity $\mathcal{T}_{\rm{s}}$ can be estimated from ${\vec{J}}_s^{\rm{in}}$ and ${\vec{J}}_s^{\rm{out}}$ (Eq.\ref{eq1}).


Figures \ref{fig:temp_dep}c and \ref{fig:temp_dep}d show the field dependence of the measured voltage $V$ for a \YIG/Pt bilayer and the \YIG/\CrO/Pt trilayer. In both samples---with and without a \CrO layer---the sign of $V$ reverses with the sign of $H$, and the shape of the $V$-$H$ curves agree with the $M$-$H$ (hysteresis) curve of the \YIG film\cite{Uchida2010, Uchida2014, Qiu2015}. This confirms that the measured voltage $V$ in the \YIG/\CrO/Pt trilayer device is induced by the thermal spin currents generated from the \YIG. 

First, we show a steep conductor-nonconductor transition for spin currents in \CrO. Figure \ref{fig:temp_dep}e shows the temperature dependence of the SSE voltage $V_{\rm{SSE}}$ ($H = 0.1$~T) for the \YIG/\CrO/Pt trilayer device. Surprisingly, the voltage exhibits an abrupt change of more than $100\times$ around 290 K. Above this temperature, a voltage with a peak of $V_{\rm{SSE}}\approx$ 500 nV appears at $T=296$~K. When $T<$ 282 K, $V_{\rm{SSE}}$ is close to the noise floor $\approx$5 nV (Fig. \ref{fig:temp_dep}e). By contrast, in the  \YIG/Pt bilayer device, $V_{\rm{SSE}}$ varies little across the same temperature range (Fig. \ref{fig:temp_dep}f)\cite{Kikkawa2015}, indicating $\vec{J}_s^{\rm in}$ is nearly constant. This equivalently means that the spin-current transmissivity $\mathcal{T}_{\rm{s}}$ of \CrO changes more than $100\times$ around 290 K, which is calculated according to Eq.\ref{eq1} and plotted in the supplementary Fig. 2c. 

We attribute the abrupt change of $V_{\rm{SSE}}$ in the \YIG/\CrO/Pt device to the change in the spin-current transmissivity $\mathcal{T}_{\rm{s}}$ of the \CrO layer, marking the transition of the \CrO layer from a spin conductor to a spin nonconductor at $T$$=$ 296 K. This critical temperature coincides with the N\'eel temperature of the \CrO thin film\cite{Tobia2008, Pati2016}, and we associate the change in spin-current transmissivity with the onset of magnetic order. We found a similar spin conductor-nonconductor transition in a spin pumping measurement for devices with the same \YIG/\CrO/Pt structure as shown in Fig. \ref{fig:temp_dep}g (also refer to supplementary
Note 1), demonstrating that the spin conductor-nonconductor transition in \CrO does not depend on the method of spin current generation. We also ruled out magnetic interface effects between the exchanged coupled \YIG and \CrO \ (such as exchange bias or spin reorientation transitions) causing the large change of $\mathcal{T}_{\rm{s}}$. Using a control sample with a 5-nm Cu layer (a nonmagnetic metal but good spin conductor) inserted between the \YIG and \CrO layers, we observed results similar to that in the \YIG/\CrO/Pt trilayer (Fig. \ref{fig:temp_dep}h, also refer to supplementary
Note 2). By measuring the $V_{\rm{SSE}}$ for a \CrO/Pt bilayer, we also confirmed that $V_{\rm{SSE}}$ comes from spin current generated in the \YIG and transmitted through the \CrO, rather than by spin current originating within \CrO (Fig. \ref{fig:temp_dep}h, also refer to supplementary
Note 2).

\begin{figure}
        \centering
        \includegraphics[width=1\linewidth]{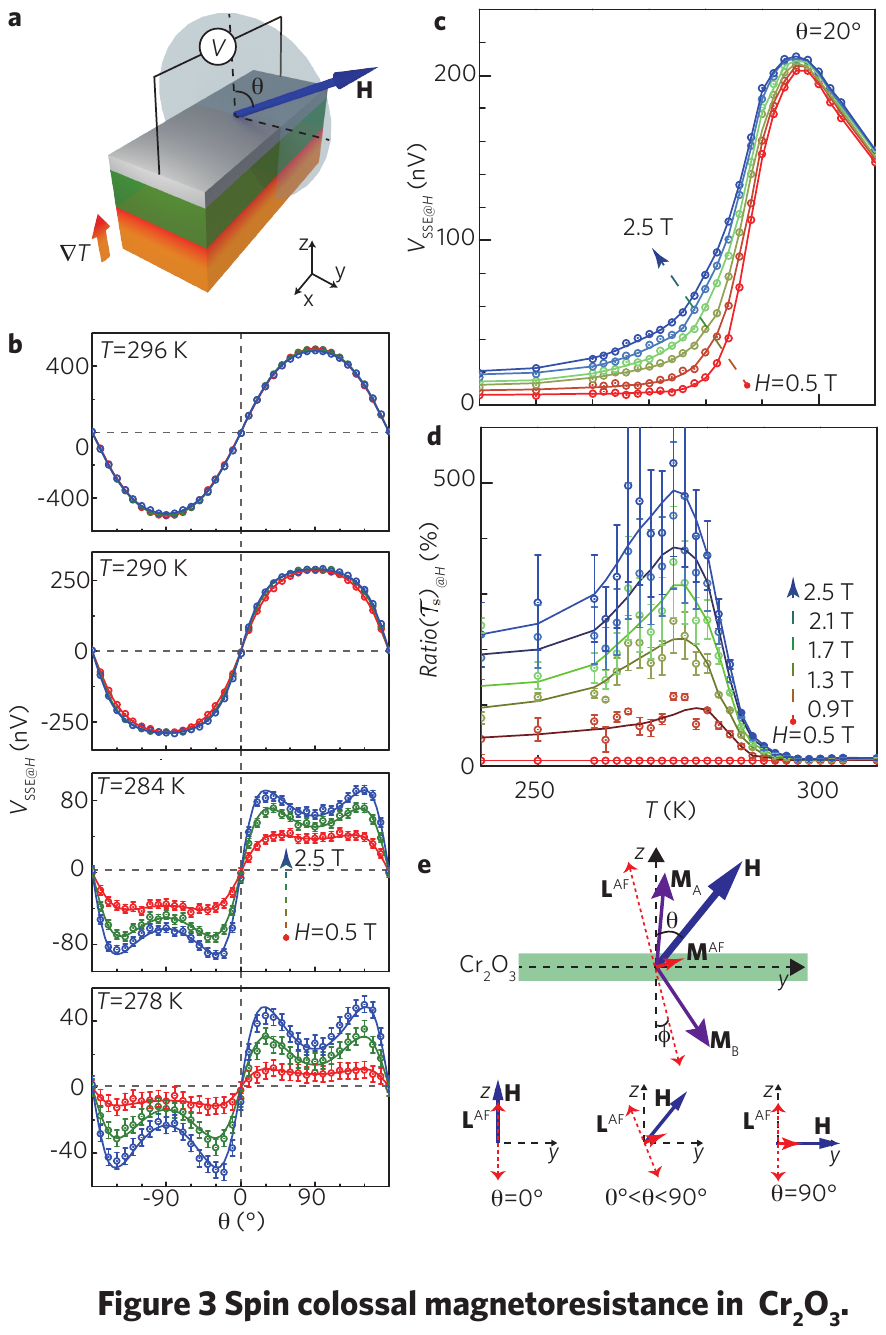}
        \caption{\textbf{Spin colossal magnetoresistance in \CrO.} 
                \textbf{a}. A schematic illustration of the out-of-plane spin Seebeck set-up for the \YIG/\CrO/Pt trilayer device. A temperature gradient, $\nabla T$, is along the $z$ direction, and an external magnetic field, $\vec{H}$, is applied in the $y$-$z$ plane. $\theta$ is the angle between $\nabla T$ and $\vec{H}$.
                \textbf{b}. $\theta$ dependencies of $V_{\rm{SSE}}$ at different temperatures for the \YIG/\CrO/Pt trilayer device with various values of $\vec{H}$. \textcolor{black}{Here, $V_{\rm{SSE}}$ refers to the spin Seebeck voltage signal detected from the Pt layer.} The solid curve is a fitting result using Eq. (\ref{eq3}). \textcolor{black}{The noise level of the voltage measurement is about 5 nV, which is smaller than the size of data points in most figures. }
                \textbf{c}. Temperature dependence of $V_{\rm{SSE}}$ for the \YIG/\CrO/Pt trilayer device at different external magnetic fields $\vec{H}$ at $\theta$=$20^{\circ}$. The solid lines are guides to the eye.
                \textbf{d}. The $\mathcal{T}_{\rm{s}}$ change ratio $Ratio(\mathcal{T}_{\rm{s}})_{@\textit{H}}$ at \textcolor{black}{an applied external magnetic field $\vec{H}$} as functions of temperature. Here, $Ratio(\mathcal{T}_{\rm{s}})_{@\textit{H}}=(V_{\rm{SSE@}\textit{H}}-V_{\rm{SSE@0.5 T}})/V_{\rm{SSE@0.5 T}}$. $\mathcal{T}_{\rm{s}}$ refers to the spin-current transmissivity in the \CrO layer. The solid lines are guides to the eye.  
                \textbf{e}. A schematic illustration of the relation between a N\'eel vector, $\vec{L}^{\rm{AF}}$, the induced magnetization $\vec{M}^{\rm{AF}}$, and the external magnetic field $\vec{H}$. Here, $\theta$ refers to the angle between $\vec{H}$ and the $z$ axis; $\phi$ refers to the angle between $\vec{L}^{\rm{AF}}$ and the easy axis $c$ of \CrO, which is along the $z$ axis in our sample. $\vec{M}_{\rm{A}}$ and $\vec{M}_{\rm{B}}$ are the magnetization vectors of the two sublattices of \CrO. The lower panel shows the relation between a N\'eel vector, $\vec{L}^{\rm{AF}}$, the induced magnetization $\vec{M}^{\rm{AF}}$, and the external magnetic field $\vec{H}$ at different values of $\theta$.
\label{fig:field_dep}}
        
\end{figure}

Having established the spin conductor-nonconductor transition, we show that the spin-current transmissivity of \CrO has an anisotropic response to magnetic fields in the critical region of the magnetic transition. The spin-current transmissivity of \CrO depends not only on the magnitude but also on the direction of the magnetic field. By using the SSE and ISHE as sources and probes of spin currents, within the critical region we measured the dependence of $V_{\rm{SSE}}$ on the magnetic field magnitude $|\vec{H}|$ and angle $\theta$ in the $z$-$y$ plane as illustrated in Fig. \ref{fig:field_dep}a. 

Figure \ref{fig:field_dep}b shows the dependence of $V_{\rm{SSE}}$  on the angle $\theta$ at different magnetic field magnitudes. At $T=$ 296 K (in the spin conducting regime), $V_{\rm{SSE}}$ shows a sinusoidal change with respect to $\theta$, the same as the relative angle between the \YIG magnetization $\vec{M}$ and ${\vec{J}}_s^{\rm{in}}$ as expected from Eq. (\ref{eq1}). The magnitude of $V_{\rm{SSE}}$ changes only slightly from $|\vec{H}|=$ 0.5 T to 2.5 T. Similar behaviour is observed for $T>$ 296 K, indicating that $\mathcal{T}_{\rm{s}}$  depends only weakly on $\theta$ or $H$ in the spin conductor regime. However, at $T<$ 296 K, $V_{\rm{SSE}}(\theta)$ starts to deviate from this dependence. As the temperature decreases further, the character of $V_{\rm{SSE}}$($\theta$) changes completely. The maximum amplitude of $V_{\rm{SSE}}$ no longer resides at $\theta$= $\pm 90^{\circ}$ but peaks four times through the rotation ($-180^{\circ}$,$180^{\circ}$). $\mathcal{T}_{\rm{s}}$ also becomes strongly dependent on $|\vec{H}|$. Thus, $\mathcal{T}_{\rm{s}}$($\theta$, $H$) depends on both $\theta$ and $|\vec{H}|$ in the critical region. 

Figure \ref{fig:field_dep}c shows the temperature dependence of $V_{\rm{SSE}}(|\vec{H}|)$  at $\theta=$20$^{\circ}$, where the $|\vec{H}|$ dependence is the most pronounced. The temperature dependence of $V_{\rm{SSE}}$ is qualitatively similar for all field strengths, featuring a sharp transition between the spin nonconductor and conductor regimes. However, the transition edge of $V_{\rm{SSE}}$ shifts to lower temperatures for stronger magnetic fields. Taking $|\vec{H}|=0.5~\mathrm{T}$ as a reference, $\sim$500\% modulation of $V_{\rm{SSE}}$ is achieved with a $2.5$~T field (Fig. \ref{fig:field_dep}d).  


Above the N\'eel temperature, the paramagnetic moments of \CrO follow the external magnetic field and spin current is carried by correlations of the paramagnetic moments as has been reported previously~\cite{Wu-PhysRevLett-2015,Wang2015a,Qiu-NatComms-2016}. 

Below the N\'eel temperature---in the ordered antiferromagnetic phase---the propagation of the spin current is in principle determined by the thermal population of magnons, the magnon mean free path and the magnon gap. However, the magnon gap is approximately 10~K\cite{Foner-PhysRev-1963}, therefore this description by itself cannot lead to the sharp transition observed at the N\'eel point. In other words, the nonconducting regime cannot be caused by magnon freezing. Rather, it is caused by the anisotropic transmissivity of the antiferromagnet in combination with the device geometry. 

Only the spin component which is parallel (or antiparallel) to the N\'eel vector can be carried by magnons\cite{Kim2014}. Below the N\'eel temperature, due to the strong uniaxial anisotropy, N\'eel vector of \CrO is pinned to the easy axis (out of plane in this work). When the YIG magnetization is in the plane of the film, the spins are polarized perpendicularly to the \CrO N\'eel vector and the spin current cannot be transmitted into the \CrO. When the YIG magnetization is out of the plane, the spin current can be transmitted but the device geometry prohibits the generation of an ISHE voltage. Furthermore, the strength of the anisotropy in \CrO is almost independent of temperature, collapsing to zero only very close to the N\'eel temperature\cite{Foner-PhysRev-1963}. Therefore, the \CrO is strongly aligned perpendicular to the plane for almost the entire temperature range and no spin current can be transmitted. This small temperature window where the anisotropy decreases corresponds with the increase in ISHE voltage. 

In the region just below the N\'eel temperature, where the anisotropy is reducing, the transmissivity can be manipulated with the applied field. The enhanced susceptibility and reduced anisotropy in this small temperature window allows the N\'eel vector to be slightly rotated, giving a finite $y$-component (in the plane) on to which the spin current is projected\cite{nagamiyaAFM, Foner-PhysRev-1963}. The field induced N\'eel vector and magnetization $y$-components of the antiferromagnet (Fig. \ref{fig:field_dep}e) are 
\begin{equation}
L_y^{\rm{AF}} \approx - \frac{M_s H^2}{2H_\mathrm{exch}H_\mathrm{ani}}  \sin2\theta;
\quad
M_y^{\rm{AF}} \approx \frac{M_s H}{H_\mathrm{exch}}\sin\theta,
\label{eq2}
\end{equation}
respectively. $M_s$ is the saturation magnetization of the antiferromagnetic sublattices, $H_\mathrm{exch}$ is the exchange field between the sublattices, and $H_\mathrm{ani}$ is the uniaxial anisotropy field. The equation is based on a zero-temperature theory but by allowing the temperature dependence of $H_\mathrm{exch}$ and $H_\mathrm{ani}$, it appears to be a good approximation even up to the N\'eel temperature. At temperatures much below the N\'eel temperature, the field required to manipulate the N\'eel vector is approximately $H\approx 6$~T~\cite{Foner-PhysRev-1963}. But, when $H_\mathrm{ani}$ drops in the transition window, much smaller fields (smaller than the spin flop field) can manipulate the N\'eel vector.

Under the assumption that spin transport is possible only for angular momentum along the N\'eel vector and in the linear dynamics regime, we phenomenologically obtain the angular dependence of the ISHE voltage as
\begin{equation}
V(\theta) = -aL_{y}^{\rm{AF}}\cos{\theta} + b M_y^{\rm{AF}} 
\label{eq3}
\end{equation}
where $a$ and $b$ are phenomenological parameters. Equations \ref{eq2} and \ref{eq3} qualitatively reproduce our experimental results for the $\theta$ dependence (Fig.~\ref{fig:field_dep}b) of the voltage (see supplementary Note 4 for details).

In summary, we report the occurrence of the spin conductor-nonconductor transition and the field induced modulation of spin-current transmissivity in \CrO, which is reminiscent of the CMR in electronics. We attribute this `colossal' modulation of spin current to the combination of the anisotropic spin current transmission of the antiferromagnet and the device geometry, which is correlated to the N\'eel vector and anisotropy of \CrO. The SCMR may also be observed in other antiferromagnetic materials in which the N\'eel vector responds to magnetic fields. It may therefore be possible to create devices which switch between the spin insulating and conducting states---but in response to a completely different stimulus. For example switching the antiferromagnet between perpendicular states electrically\cite{Wadley2016}.



\begin{thebibliography}{10}
	\expandafter\ifx\csname url\endcsname\relax
	\def\url#1{\texttt{#1}}\fi
	\expandafter\ifx\csname urlprefix\endcsname\relax\def\urlprefix{URL }\fi
	\providecommand{\bibinfo}[2]{#2}
	\providecommand{\eprint}[2][]{\url{#2}}
	
	\bibitem{Ramirez1997}
	\bibinfo{author}{Ramirez, A.~P.}
	\newblock \bibinfo{title}{{Colossal magnetoresistance}}.
	\newblock \emph{\bibinfo{journal}{Journal of Physics: Condensed Matter}}
	\textbf{\bibinfo{volume}{9}}, \bibinfo{pages}{8171--8199} (\bibinfo{year}{1997}).
	
	\bibitem{Tokura2006}
	\bibinfo{author}{Tokura, Y.}
	\newblock \bibinfo{title}{{Critical features of colossal magnetoresistive
			manganites}}.
	\newblock \emph{\bibinfo{journal}{Reports on Progress in Physics}}
	\textbf{\bibinfo{volume}{69}}, \bibinfo{pages}{797--851} (\bibinfo{year}{2006}).
	
	\bibitem{Zutic2004}
	\bibinfo{author}{{\v{Z}}uti{\'{c}}, I.} \& \bibinfo{author}{{Das Sarma}, S.}
	\newblock \bibinfo{title}{{Spintronics: Fundamentals and applications}}.
	\newblock \emph{\bibinfo{journal}{Reviews of Modern Physics}}
	\textbf{\bibinfo{volume}{76}}, \bibinfo{pages}{323--410}
	(\bibinfo{year}{2004}).
	
	\bibitem{Maekawa2006}
	\bibinfo{author}{Maekawa, S.}
	\newblock \emph{\bibinfo{title}{{Concepts in Spin Electronics}}}
	(\bibinfo{publisher}{Oxford Univ. Press}, \bibinfo{year}{2006}).
	
	\bibitem{Kajiwara2010}
	\bibinfo{author}{Kajiwara, Y.} \emph{et~al.}
	\newblock \bibinfo{title}{{Transmission of electrical signals by spin-wave
			interconversion in a magnetic insulator.}}
	\newblock \emph{\bibinfo{journal}{Nature}} \textbf{\bibinfo{volume}{464}},
	\bibinfo{pages}{262--266} (\bibinfo{year}{2010}).
	
	\bibitem{Uchida-Nature-2008}
	\bibinfo{author}{Uchida, K.} \emph{et~al.}
	\newblock \bibinfo{title}{Observation of the spin seebeck effect}.
	\newblock \emph{\bibinfo{journal}{Nature}} \textbf{\bibinfo{volume}{455}},
	\bibinfo{pages}{778--781} (\bibinfo{year}{2008}).
	
	\bibitem{Takei-PhysRevB-2014}
	\bibinfo{author}{Takei, S.}, \bibinfo{author}{Halperin, B.~I.},
	\bibinfo{author}{Yacoby, A.} \& \bibinfo{author}{Tserkovnyak, Y.}
	\newblock \bibinfo{title}{Superfluid spin transport through antiferromagnetic
		insulators}.
	\newblock \emph{\bibinfo{journal}{Phys. Rev. B}} \textbf{\bibinfo{volume}{90}},
	\bibinfo{pages}{094408} (\bibinfo{year}{2014}).
	
	\bibitem{Cornelissen-NatPhys-2015}
	\bibinfo{author}{Cornelissen, L.}, \bibinfo{author}{Liu, J.},
	\bibinfo{author}{Duine, R.}, \bibinfo{author}{Ben~Youssef, J.} \&
	\bibinfo{author}{Van~Wees, B.}
	\newblock \bibinfo{title}{Long-distance transport of magnon spin information in
		a magnetic insulator at room temperature}.
	\newblock \emph{\bibinfo{journal}{Nature Physics}}
	\textbf{\bibinfo{volume}{11}}, \bibinfo{pages}{1022--1026}
	(\bibinfo{year}{2015}).
	
	\bibitem{Qiu-NatComms-2016}
	\bibinfo{author}{Qiu, Z.} \emph{et~al.}
	\newblock \bibinfo{title}{Spin-current probe for phase transition in an
		insulator}.
	\newblock \emph{\bibinfo{journal}{Nature communications}}
	\textbf{\bibinfo{volume}{7}}, \bibinfo{pages}{12670} (\bibinfo{year}{2016}).
	
	\bibitem{Tserkovnyak-NatNano-2013}
	\bibinfo{author}{Tserkovnyak, Y.}
	\newblock \bibinfo{title}{Spintronics: An insulator-based transistor}.
	\newblock \emph{\bibinfo{journal}{Nature nanotechnology}}
	\textbf{\bibinfo{volume}{8}}, \bibinfo{pages}{706--707}
	(\bibinfo{year}{2013}).
	
	\bibitem{Brockhouse-JChemPhys-1953}
	\bibinfo{author}{Brockhouse, B.~N.}
	\newblock \bibinfo{title}{Antiferromagnetic structure in {Cr$_{2}$O$_3$}}.
	\newblock \emph{\bibinfo{journal}{The Journal of Chemical Physics}}
	\textbf{\bibinfo{volume}{21}}, \bibinfo{pages}{961--962}
	(\bibinfo{year}{1953}).
	
	\bibitem{Foner-PhysRev-1963}
	\bibinfo{author}{Foner, S.}
	\newblock \bibinfo{title}{High-field antiferromagnetic resonance in
		{Cr$_{2}$O$_3$}}.
	\newblock \emph{\bibinfo{journal}{Physical Review}}
	\textbf{\bibinfo{volume}{130}}, \bibinfo{pages}{183--197}
	(\bibinfo{year}{1963}).
	
	\bibitem{nagamiyaAFM}
	\bibinfo{author}{Nagamiya, T.}, \bibinfo{author}{Yosida, K.} \&
	\bibinfo{author}{Kubo, R.}
	\newblock \bibinfo{title}{Antiferromagnetism}.
	\newblock \emph{\bibinfo{journal}{Advances in Physics}}
	\textbf{\bibinfo{volume}{4}}, \bibinfo{pages}{1--112} (\bibinfo{year}{1955}).
	
	\bibitem{Xiao-PhysRevB-2010}
	\bibinfo{author}{Xiao, J.} \emph{et~al.}
	\newblock \bibinfo{title}{Theory of magnon-driven spin seebeck effect}.
	\newblock \emph{\bibinfo{journal}{Physical Review B}}
	\textbf{\bibinfo{volume}{81}}, \bibinfo{pages}{214418}
	(\bibinfo{year}{2010}).
	
	\bibitem{Saitoh2006}
	\bibinfo{author}{Saitoh, E.}, \bibinfo{author}{Ueda, M.},
	\bibinfo{author}{Miyajima, H.} \& \bibinfo{author}{Tatara, G.}
	\newblock \bibinfo{title}{{Conversion of spin current into charge current at
			room temperature: Inverse spin-Hall effect}}.
	\newblock \emph{\bibinfo{journal}{Applied Physics Letters}}
	\textbf{\bibinfo{volume}{88}}, \bibinfo{pages}{182509}
	(\bibinfo{year}{2006}).
	
	\bibitem{Wang2015a}
	\bibinfo{author}{Wang, H.}, \bibinfo{author}{Du, C.}, \bibinfo{author}{Hammel,
		P.~C.} \& \bibinfo{author}{Yang, F.}
	\newblock \bibinfo{title}{{Spin transport in antiferromagnetic insulators
			mediated by magnetic correlations}}.
	\newblock \emph{\bibinfo{journal}{Physical Review B}}
	\textbf{\bibinfo{volume}{91}}, \bibinfo{pages}{220410(R)}
	(\bibinfo{year}{2015}).
	
	\bibitem{Moriyama2015}
	\bibinfo{author}{Moriyama, T.} \emph{et~al.}
	\newblock \bibinfo{title}{{Anti-damping spin transfer torque through epitaxial
			nickel oxide}}.
	\newblock \emph{\bibinfo{journal}{Applied Physics Letters}}
	\textbf{\bibinfo{volume}{106}}, \bibinfo{pages}{162406}
	(\bibinfo{year}{2015}).
	
	\bibitem{Uchida2010}
	\bibinfo{author}{Uchida, K.} \emph{et~al.}
	\newblock \bibinfo{title}{{Spin Seebeck insulator}}.
	\newblock \emph{\bibinfo{journal}{Nature Materials}}
	\textbf{\bibinfo{volume}{9}}, \bibinfo{pages}{894--897}
	(\bibinfo{year}{2010}).
	
	\bibitem{Uchida2014}
	\bibinfo{author}{Uchida, K.} \emph{et~al.}
	\newblock \bibinfo{title}{{Longitudinal spin Seebeck effect: from fundamentals
			to applications.}}
	\newblock \emph{\bibinfo{journal}{Journal of physics. Condensed matter : an
			Institute of Physics journal}} \textbf{\bibinfo{volume}{26}},
	\bibinfo{pages}{343202} (\bibinfo{year}{2014}).
	
	\bibitem{Qiu2015}
	\bibinfo{author}{Qiu, Z.}, \bibinfo{author}{Hou, D.}, \bibinfo{author}{Uchida,
		K.} \& \bibinfo{author}{Saitoh, E.}
	\newblock \bibinfo{title}{{Influence of interface condition on spin-Seebeck
			effects}}.
	\newblock \emph{\bibinfo{journal}{Journal of Physics D: Applied Physics}}
	\textbf{\bibinfo{volume}{48}}, \bibinfo{pages}{164013}
	(\bibinfo{year}{2015}).
	
	\bibitem{Kikkawa2015}
	\bibinfo{author}{Kikkawa, T.} \emph{et~al.}
	\newblock \bibinfo{title}{Critical suppression of spin seebeck effect by
		magnetic fields}.
	\newblock \emph{\bibinfo{journal}{Physical Review B}}
	\textbf{\bibinfo{volume}{92}}, \bibinfo{pages}{064413}
	(\bibinfo{year}{2015}).
	
	\bibitem{Tobia2008}
	\bibinfo{author}{Tobia, D.}, \bibinfo{author}{Winkler, E.},
	\bibinfo{author}{Zysler, R.~D.}, \bibinfo{author}{Granada, M.} \&
	\bibinfo{author}{Troiani, H.~E.}
	\newblock \bibinfo{title}{{Size dependence of the magnetic properties of
			antiferromagnetic Cr$_2$O$_3$ nanoparticles}}.
	\newblock \emph{\bibinfo{journal}{Physical Review B}}
	\textbf{\bibinfo{volume}{78}}, \bibinfo{pages}{104412}
	(\bibinfo{year}{2008}).
	
	\bibitem{Pati2016}
	\bibinfo{author}{Pati, S.~P.} \emph{et~al.}
	\newblock \bibinfo{title}{{Finite-size scaling effect on N\'eel temperature of
			antiferromagnetic Cr$_2$O$_3$ (0001) films in exchange-coupled
			heterostructures}}.
	\newblock \emph{\bibinfo{journal}{Physical Review B}}
	\textbf{\bibinfo{volume}{94}}, \bibinfo{pages}{224417}
	(\bibinfo{year}{2016}).
	
	\bibitem{Wu-PhysRevLett-2015}
	\bibinfo{author}{Wu, S.~M.}, \bibinfo{author}{Pearson, J.~E.} \&
	\bibinfo{author}{Bhattacharya, A.}
	\newblock \bibinfo{title}{Paramagnetic spin seebeck effect}.
	\newblock \emph{\bibinfo{journal}{Phys. Rev. Lett.}}
	\textbf{\bibinfo{volume}{114}}, \bibinfo{pages}{186602}
	(\bibinfo{year}{2015}).
	
	\bibitem{Kim2014}
	\bibinfo{author}{Kim, S.~K.}, \bibinfo{author}{Tserkovnyak, Y.} \&
	\bibinfo{author}{Tchernyshyov, O.}
	\newblock \bibinfo{title}{Propulsion of a domain wall in an antiferromagnet by
		magnons}.
	\newblock \emph{\bibinfo{journal}{Physical Review B}}
	\textbf{\bibinfo{volume}{90}}, \bibinfo{pages}{104406} (\bibinfo{year}{2014}).
	
	\bibitem{Wadley2016}
	\bibinfo{author}{Wadley, P.} \emph{et~al.}
	\newblock \bibinfo{title}{Electrical switching of an antiferromagnet}.
	\newblock \emph{\bibinfo{journal}{Science}} \textbf{\bibinfo{volume}{351}},
	\bibinfo{pages}{587--590} (\bibinfo{year}{2016}).
	
\end{thebibliography}

\begin{thebibliography}{10}
	\expandafter\ifx\csname url\endcsname\relax
	\def\url#1{\texttt{#1}}\fi
	\expandafter\ifx\csname urlprefix\endcsname\relax\def\urlprefix{URL }\fi
	\providecommand{\bibinfo}[2]{#2}
	\providecommand{\eprint}[2][]{\url{#2}}
	
\bibitem[27]{Vasyuchka2014}
\bibinfo{author}{Vasyuchka, V.~I.}
\newblock \bibinfo{title}{{Microwave-induced spin currents in
		ferromagnetic-insulator|normal-metal bilayer system}}.
\newblock \emph{\bibinfo{journal}{Applied Physics Letters}}
\textbf{\bibinfo{volume}{105}}, \bibinfo{pages}{092404}
(\bibinfo{year}{2014}).

\end{thebibliography}


\begin{addendum}
 \item [Acknowledgements] This work was supported by JST-ERATO `Spin Quantum Rectification', JST-PRESTO `Phase Interfaces for Highly Efficient Energy Utilization', Grant-in-Aid for Scientific Research on Innovative Area, `Nano Spin Conversion Science' (26103005 and 26103006), Grant-in-Aid for Scientific Research (S) (25220910), Grant-in-Aid for Scientific Research (A) (25247056 and 15H02012), Grant-in-Aid for Challenging Exploratory Research (26600067), Grant-in-Aid for Research Activity Start-up (25889003), and World Premier International Research Center Initiative (WPI), all from MEXT, Japan, ZQ acknowledges support from "the Fundamental Research Funds for the Central Universities (DUT17RC(3)073)". DH acknowledges support from Grant-in-Aid for young scientists (B) (JP17K14331), JB acknowledges supports from the Graduate Program in Spintronics, Tohoku University, and Grand-in-Aid for Young Scientists (B) (17K14102). KY and OG acknowledge support from Humboldt Foundation and EU ERC Advanced Grant No. 268066. KY acknowledges the Transregional Collaborative Research Center (SFB/TRR) 173 SPIN+X and DAAD project "MaHoJeRo". O.G. acknowledge the EU FET Open RIA Grant no. 766566 and  the DFG (project SHARP 397322108).
 
 \item[Author contributions]Z.Q. and D.H. designed the experiment, Z.Q. fabricated the samples and collected all of the data. Z.Q., D.H., J.B. and K.Y. analyzed the data. J.B., K.Y. and O.G. contribute theoretical discussions. E.S. supervised this study. All the authors discussed the results and prepared the manuscript.
 \item[Competing Interests] The authors declare that they have no competing financial interests.
 \item[Correspondence] Correspondence and requests for materials should be addressed to D.H. (email: dazhi.hou@imr.tohoku.ac.jp).
\end{addendum}

\begin{methods}
	\subsection{Preparation of \YIG/\CrO/Pt samples.} 
	We grew a 3 $\mu$m-thick single-crystalline \YIG film on a (111) Gd$_3$Ga$_5$O$_{12}$ wafer by liquid phase epitaxy method at 1203 K in PbO-B$_2$O$_3$ based flux. We cut a single wafer into 1.5 mm$\times$3 mm in size. A 12 nm-thick \CrO film is grown on  top of the \YIG film by  pulsed laser deposition  at 673 K and subsequently annealed at 1073 K for 30 min to obtain continuous films and improve the crystallinity. 10 nm-thick Pt films were then grown on the top of the \CrO by RF magnetron sputtering.
	
	\subsection{Sample characterization.} 
	Crystallographic characterization for the samples was carried out by X-ray diffractometry and transmission electron microscopy (TEM). The obtained TEM image shows that the \YIG\ film is of a single-crystal structure, and the easy axis ($c$-axis) of the hexagonal \CrO grown on the top of the \YIG is along the out-of-plane direction $z$ (Fig. \ref{fig:temp_dep}b).

	\subsection{Spin Seebeck experimental set-up.}
	We performed spin Seebeck measurements in a vector magnet system (Oxford instruments inc). We set the samples on a wave guide and heated the Pt layer by using a pulsed microwave\cite{Vasyuchka2014} (8 GHz, 1 W), which creates a temperature gradient as shown in Fig. \ref{fig:temp_dep}a. We measured the voltage signal between the ends of the Pt layer using a lock-in amplifier. 
	
\end{methods}

\textbf{Data availability}

The data that support the findings of this study are available from the authors on reasonable request, see author contributions for specific data sets.

\end{document}